\documentclass[12pt,prd,showpacs,nofootinbib]{revtex4}
\usepackage{epsf,graphicx}
\usepackage{amsmath,amsfonts,amssymb,latexsym}

\newcommand{\bls}[1]{\renewcommand{\baselinestretch}{#1}}

\def\noi{\noindent}

\def\cm{\hspace*{1cm}}

\def\Jl#1#2{#1 {\bf #2},\ }

\def\ApJ#1 {\Jl{Astroph. J.}{#1}}
\def\CQG#1 {\Jl{Class. Quantum Grav.}{#1}}
\def\DAN#1 {\Jl{Dokl. AN SSSR}{#1}}
\def\GC#1 {\Jl{Grav. Cosmol.}{#1}}
\def\GRG#1 {\Jl{Gen. Rel. Grav.}{#1}}
\def\JETF#1 {\Jl{Zh. Eksp. Teor. Fiz.}{#1}}
\def\JETP#1 {\Jl{Sov. Phys. JETP}{#1}}
\def\JHEP#1 {\Jl{JHEP}{#1}}
\def\JMP#1 {\Jl{J. Math. Phys.}{#1}}
\def\NPB#1 {\Jl{Nucl. Phys. B}{#1}}
\def\NP#1 {\Jl{Nucl. Phys.}{#1}}
\def\PLA#1 {\Jl{Phys. Lett. A}{#1}}
\def\PLB#1 {\Jl{Phys. Lett. B}{#1}}
\def\PRD#1 {\Jl{Phys. Rev. D}{#1}}
\def\PRL#1 {\Jl{Phys. Rev. Lett.}{#1}}

\def\al{&}
\def\lal{&&{}}
\def\eq{Eq.\,}
\def\eqs{Eqs.\,}
\def\beq{\begin{equation}}
\def\eeq{\end{equation}}
\def\bear{\begin{eqnarray}}
\def\bearr{\begin{eqnarray} \lal}
\def\ear{\end{eqnarray}}
\def\earn{\nonumber \end{eqnarray}}
\def\nn{\nonumber\\ {}}

\def\nnn{\nonumber\\ \lal }
\def\nnnv{\nonumber\\[5pt] \lal }
\def\yy{\\[5pt] {}}

\def\eql{\al =\al}

\def\tst{\textstyle}

\def\fract#1#2{{\tst\frac{#1}{#2}}}

\def\half{{\fract{1}{2}}}

\def\e{{\,\rm e}}
\def\d{\partial}

\def\sign{\mathop{\rm sign}\nolimits}
\def\diag{\mathop{\rm diag}\nolimits}

\def\const{{\rm const}}
\def\eps{\varepsilon}

\def\then{\ \Rightarrow\ }

\def\mn{_{\mu\nu}}
\def\mN{_{\mu}^{\nu}}
\def\MN{^{\mu\nu}}

\def\rh{r_{\rm hor}}
\def\xh{x_{\rm hor}}

\def\N{{\mathbb N}}
\def\R{{\mathbb R}}

\def\asflat{asymptotically flat}
\def\sph{spherically symmetric}
\def\ssph{static, spherically symmetric}
\def\bh{black hole}
\def\bhs{black holes}

\def\whs{wormholes}
\def\cy{cylindrical}

\def\RN{Reiss\-ner-Nord\-str\"om}
\bls{1.06}
\begin{document}
\thispagestyle{empty}

\title{Dilaton gravity, (quasi-) black holes, and scalar charge}

\author{K.A. Bronnikov}
\affiliation{Center for Gravitation and Fundamental Metrology,
	VNIIMS, 46 Ozyornaya Street, Moscow 119361, Russia, and\\
	Institute of Gravitation and Cosmology,
	PFUR, 6 Miklukho-Maklaya Street, Moscow 117198, Russia, and\\
	I. Kant Baltic Federal University, Al. Nevsky St. 14,
	Kaliningrad 236041, Russia}
	\email{kb20@yandex.ru}

\author{J.C. Fabris and \fbox{R. Silveira}}
\affiliation{Universidade Federal do Esp\'{\i}rito Santo, Departamento de F\'{\i}sica,
	Av. Fernando Ferrari 514, Campus de Goiabeiras, CEP 29075-910, Vit\'oria, ES, Brazil}
	\email{fabris@pq.cnpq.br}

\author{O.B. Zaslavskii}
\affiliation{Department of Physics and Technology, Kharkov V.N. Karazin National
	University, 4 Svoboda Square, Kharkov, 61077, Ukraine, and\\
	Institute of Mathematics and Mechanics, Kazan Federal University,
	18 Kremlyovskaya Street, Kazan 420008, Russia}
	\email{zaslav@ukr.net}

\begin{abstract}
  Electrically charged dust is considered in the framework of
  Einstein-Maxwell-dilaton gravity with a Lagrangian containing the
  interaction term $P(\chi)F_{\mu\nu}F^{\mu\nu}$, where $P(\chi)$ is an
  arbitrary function of the dilaton scalar field $\chi$, which can be
  normal or phantom. Without assumption of spatial symmetry, we show that
  static configurations exist for arbitrary functions $g_{00}
  = \exp(2\gamma(x^{i}))$ ($i=1,2,3$) and $\chi =\chi(\gamma)$. If
  $\chi = \const$, the classical Majumdar-Papapetrou (MP) system is
  restored. We discuss solutions that represent black holes (BHs) and
  quasi-black holes (QBHs), deduce some general results and confirm them
  by examples. In particular, we analyze configurations with spherical
  and cylindrical symmetries. It turns out that cylindrical BHs and QBHs
  cannot exist without negative energy density somewhere in space.
  However, in general, BHs and QBHs can be phantom-free, that is, can exist
  with everywhere nonnegative energy densities of matter, scalar and
  electromagnetic fields.
\end{abstract}

\maketitle

\section{Introduction}

  Studies of self-gravitating systems with spherical symmetry are one of the
  most important applications of gravity theories, above all because they
  are directly related to one of the most notable objects in nature: the
  stars. General Relativity (GR) modifies the usual predictions of the
  Newtonian theory for static, spherically symmetric systems in many
  different ways, and at the same time makes possible the emergence of a new
  class of structures, black holes (BHs), collapsed objects that can
  represent the final fate of supermassive stars. The crucial concept
  characterizing BHs is the existence of an event horizon, a hypersurface
  isolating the internal region of a collapsed object from the external
  region with distant observers living in almost flat space-time.

  The vacuum static, spherically symmetric solution of GR, known as the
  Schwarzschild solution, describes the simplest BH. Its generalizations are
  the charged static black hole represented by the Reissner-Nordtstr\"om
  (RN) solution, rotating (Kerr) and charged rotating (Kerr-Newman) black
  holes \cite{h-e, fro-nov}. In the non-vacuum case, one important type of
  configurations are those of static charged dust, represented by the
  Majumdar-Papapetrou (MP) solution \cite{majumdar,papa}:  it represents an
  equilibrium between gravitational attraction and electric repulsion.

  An important feature of the MP system is that it actually does not need
  spherical or any other spatial symmetry for its existence: the static
  equilibrium between gravity and electric forces allows for any spatial
  shape of a charged dust cloud provided the mass to charge density ratio
  takes everywhere the proper value: $\rho_e/\rho_m = \pm 1$ in natural units
  ($c = G = 1$, $c$ being the speed of light and $G$ the Newtonian
  gravitational constant).

  The MP systems have been revived in quite a new context connected with the
  so-called quasi-black holes (QBHs) [5--12]. 
  These are
  systems on the threshold of forming a black hole horizon but the horizon
  does not form. This becomes possible since equilibrium between
  gravitational attraction and electrostatic repulsion exists for any size
  of the object close to the horizon radius. From the viewpoint of a distant
  observer, the system looks indistinguishable from a true BH, although in
  the vicinity of this would-be horizon their properties are very different.

  All these examples are implemented in the pure GR context, using at most
  the electromagnetic field as an external source. Less known systems are
  those which include scalar fields. One initial reason for ignoring scalar
  fields were doubts on the existence of fundamental scalar particles in
  nature. The recently announced discovery of the Higgs boson may invalidate
  these doubts. Anyway, BHs with scalar fields have been considered many
  times in the literature since scalar fields are the simplest possible
  sources of the gravitational field.

  Scalar fields minimally coupled to gravity were considered for the
  first time, to our knowledge, by Fisher \cite{fish} who found a \ssph\
  solution to the Einstein-massless scalar field equations. A counterpart of
  Fisher's solution for massless scalar fields with a wrong sign of the
  kinetic energy (the so-called phantom scalar) was found by Bergmann and
  Leipnik \cite{pha-57}, and both solutions were repeatedly re-discovered
  afterwards. Later, a general study of spherically symmetric scalar +
  gravity systems, including a possible non-minimal coupling (hence in the
  context of a general scalar-tensor theory of gravity rather than pure GR),
  was undertaken in \cite{br73}\footnote
  	{This paper, entitled ``Scalar-tensor theory and scalar charge'',
	was published 40 years ago. The title of the present paper is
	chosen deliberately to mark this date.},
  where some nontrivial new configurations corresponding to BH and wormhole
  structures were identified. Among special cases thereof are neutral and
  charged BHs \cite{boch} and \whs\ with a conformally coupled scalar field
  as well as the so-called Ellis \whs\ (described by the simplest case of
  the Bergmann-Leipnik solution \cite{pha-57}) and \whs\ in the Brans-Dicke
  scalar-tensor theory \cite{BD} in the case where the coupling constant
  $\omega$ is smaller than $-3/2$. These results were somehow extended later
  by identifying the so-called cold BHs, i.e., BHs containing scalar
  charges, exhibiting zero surface gravity and infinite horizon surface
  areas \cite{cold1,cold2,cold-q}. In general, to obtain such cold BHs, the scalar
  field must have negative kinetic energy, that is, it must be phantom and
  violate the standard energy conditions.

  The emergence of string theories as a possible candidate for a unified
  theory of all interactions has led to new ingredients with important
  consequences for the studies of structures like BHs. String theory in its
  effective low-energy formulation predicts the existence of scalar and
  gauge fields with a nontrivial coupling between them. More precisely, the
  string effective action contains a scalar field, the dilaton, nonmiminally
  coupled to gravity. If the electromagnetic field is kept in the
  Neveu-Schwarz sector of the tree-level string action, re-expressed in the
  Einstein frame, the coupling between the dilatonic and electromagnetic
  fields appears in an exponential form, as $\e^{2\lambda\chi} F\mn F\MN$,
  where $\chi$ is the dilaton. However, taking into account the loop
  expansion of string theory, one can obtain other (and hard-to-identify)
  coupling functions $P(\chi)$ instead of $\e^{2\lambda\chi}$ \cite{dam1,
  dam2}. This motivates studies of the Einstein-Maxwell-dilaton (EMD) system
  with an arbitrary function $P(\chi)$.

  Concerning exact static, spherically symmetric solutions for the EMD
  system, to the best of our knowledge, the solutions for $P(\chi) =
  \e^{2\lambda\chi}$ were obtained for the first time in \cite{BSh-77,BSh-77e},
  where also a family of what we now know as ``dilatonic \bhs'' \cite{dbh1,
  dbh2, dbh3} was singled out. Later on, the EMD system with an arbitrary
  function $P(\chi)$ was considered \cite{BMsh-78,BMsh-78e,ann-79} in attempts 
  to obtain a nonsingular, classical, purely field particle model. The
  conditions were revealed under which a regular center in the EMD system is
  possible, and an explicit example of such a regular model was built.

  More general studies, inspired by achievements in various field limits of
  supergravity and superstring theories, were undertaken in the 90s and
  involved extensions of the EMD system to space-times of arbitrary
  dimensions with products of multiple factor spaces and, as material
  sources of gravity, a number of scalar dilatonic fields ($\chi^a$) and a
  number of antisymmetric forms of different ranks ($F_s$), of which the
  electromagnetic field $F\mn$ is the simplest representative, see
  \cite{bra-r1, bra-r2} for reviews. Various classes of exact solutions were
  found, among them many \ssph\ ones (see, e.g., \cite{kb95, bra1, bra2, bra3}
  and references therein), of which the most well-known are the so-called
  ``black branes'' with different configurations of extra factor spaces. It
  should be noted that in all these studies the interaction between scalar
  and antisymmetric form fields was assumed in the exponential form, like
  $F_s^2 \exp (\lambda_{sa}\chi^a)$, where $\lambda_{sa}$ is a matrix with
  constant real elements.

  Returning to four dimensions with a single scalar interacting
  exponentially with a single electromagnetic field, we can say that the
  most complete study of this \ssph\ EMD system was carried out in
  \cite{clem1} where the initial action was not restricted to only a
  string-inspired coupling and included the possibility of a phantom
  behavior of both scalar and Maxwell fields. Many new BH structures were
  revealed, asymptotically flat or not, but all of them required the
  existence of phantom fields, either in the scalar or in the
  electromagnetic sector (or in both). All these new structures have
  properties similar to cold black holes in the sense that their surface
  gravity is zero. (For a study of a perfect-fluid source of vacuum EMD
  fields see \cite{yaz}).

  Such an extremal nature of these phantom EMD BHs leads to a speculation
  that structures like QBHs, previously based on the Majumdar-Papapetrou
  solution of GR, could be implemented for such more general phantom BHs:
  since in all these cases the scalar field is phantom, it is possible to
  conjecture on a QBH configuration supported by an attractive or repulsive
  effect of a scalar field in addition to electric repulsion. This idea
  has been initially considered in a PhD thesis of one of the co-authors of
  the present work, Robson  Silveira, who died in 2009 before completing the
  proposed study. Before his death, he was able to obtain some initial
  results indicating that such scalar QBHs are really possible and designing
  some of their main properties. In particular, he found QBHs supported by
  only a phantom scalar field. The goal of the present paper is
  to report on a more general analysis completing his findings.

  We include into consideration all ingredients which were earlier discussed
  separately: matter (electrically and scalarly charged dust), an
  electromagnetic field and a scalar field interacting with it in a
  dilatonic manner. We make some general observations on possible equilibrium
  configurations (to be called dilatonic MP, or DMP systems) and pay special
  attention to BHs and QBHs supported by certain electric and scalar charge
  distributions. In particular, we try to find phantom-free configurations,
  i.e., those able to exist with positive-definite energy densities of matter
  and both fields. Some of the results have been already briefly presented
  in our previous paper \cite{we-14}.

  The paper is organized as follows. Section II presents the general EMD
  field equations, Section III specializes them for static systems to be
  considered, Section IV describes their two simple special cases, one of
  them being that of MP systems, the other its purely scalar counterpart. In
  Section V we make some observations about general EMD systems without
  symmetry and determine, in particular, the necessity of a nonzero scalar
  charge density for their existence. Section VI is devoted to general
  properties of \sph\ systems, including the conditions at a regular center,
  at a horizon, and at flat spatial infinity. We also obtain a balance
  condition between the mass and integral electric and scalar charges,
  applicable to any \asflat\ systems, not necessarily spherical ones. The
  general properties of \sph\ BHs and QBHs are also described, including
  limiting transitions between them. In Section VII we give some examples of
  BHs and QBHs, showing, in particular, that both are possible without
  invoking phantom matter or fields. Section VIII deals with the properties
  of cylindrically symmetric configurations. Lastly, Section IX summarizes
  and discusses the results.

\section {Basic equations}

  Consider the Lagrangian (putting $c=G=1$)
\beq    \label{L}
      L = \frac{1}{16\pi}\biggl[
      		R + 2\eps (\d\chi)^2 - F^2 P(\chi)\biggr]
					+ L_m + A_\mu j^\mu + J\chi,
\eeq
  where $\eps = \pm 1$ ($\eps=1$ for a normal scalar field $\chi$),
  $L_m$ is the Lagrangian of matter,
  $J$ is the scalar charge density,
  $F^2\equiv F^{\alpha\beta}F_{\alpha\beta}$
  ($F\mn = \d_\mu A_\nu - \d_\nu A_\mu$, the electromagnetic field),
  $j^\mu = \rho_e u^\mu$ is the 4-current, $u^\mu$ is the 4-velocity.
  For generality, and also to provide correspondence with \cite{clem1,clem2},
  we do not fix the sign of $P(\chi)$.

  The corresponding field equations are
\bear 					                  \label{eq-s}
      4\eps \Box \chi + F^2 P_\chi \eql 16\pi J,
\yy                                  			  \label{eq-F}
      \nabla_\alpha(P(\chi) F^{\alpha\mu}) \eql -4\pi j^\mu,
\yy                                             	  \label{EE}
	G\mN \eql -8\pi T\mN,
\ear
  where $P_\chi := dP/d\chi$ and $T\mN$ is the summed stress-energy tensor
  (SET) of matter and the fields:
\bear
	T\mN \eql T\mN{}_{(s)} + T\mN{}_{(e)} + T\mN{}_{(m)}, \label{SET}
\yy
	8\pi T\mN{}_{(s)} \eql \eps[2\d_\mu\chi\d^\nu\chi
			 - \delta\mN (\d\chi)^2],    \label{SET-s}
\yy
	8\pi T\mN{}_{(e)} \eql P(\chi)               \label{SET-e}
		[-2 F_{\mu\alpha}F^{\nu\alpha} + \half \delta\mN F^2],
\yy
	T\mN{}_{(m)} \eql \rho_m u_\mu u^\nu.           \label{SET-m}
\ear

  The condition $\nabla_\nu T\mN = 0$ leads to the equations of motion for
  dust particles (using \eqs (\ref{eq-s}) and (\ref{eq-F}) to eliminate
  the second-order derivatives)
\beq                                                     \label{cons}
       u_\mu \nabla_\alpha(\rho_m u^\alpha)
       		+ \rho_m u^\alpha \nabla_\alpha u_\mu
       			- F_{\mu \alpha} j^{\alpha} + J \chi_{,\mu}=0.
\eeq
  Its contraction with $u^\mu$ leads to the generalized continuity equation
\beq
	\nabla_\alpha (\rho_m u^{\alpha})                \label{contin}
			+ J \chi_{,\alpha} u^\alpha = 0.
\eeq
  With (\ref{contin}), \eq (\ref{cons}) takes the form
\beq                                                     \label{motion}
       \rho_m u^\alpha \nabla_\alpha u_\mu
       		- F_{\mu \alpha} j^{\alpha}
	   	     + J (\chi_{,\mu} - u_\mu u^\alpha \chi_{,\alpha})=0.
\eeq

\section {Static equilibrium}

  The assumptions of static equilibrium are: the metric
\bear                                                       \label{ds}
      ds^2 = \e^{2\gamma} dt^2 - \e^{-2\gamma} h_{ik} dx^i dx^k,
\ear
  and only $F_{0i} = -F_{i0} = \phi_i$ are nonzero among $F\mn$ (only the
  electric field is present);
  $\gamma$, $h_{ik}$, $\phi$, $\chi$ are functions of $x^i$, $i = 1,2,3$.
  We use the notations $\gamma_i = \d_i\gamma$, $\phi_i =\d_i \phi$ etc.
  The spatial indices are raised and lowered with the metric $h_{ik}$ and
  its inverse $h^{ik}$, and the 4-velocity is $u^\mu = \delta^\mu_0
  \e^{-\gamma}$.

  \eqs (\ref{eq-s}) and (\ref{eq-F}) take the form
\bear                                                   \label{eq-s1}
      	2\eps \e^{2\gamma} \Delta \chi
			+ P_\chi \phi_i \phi^i \eql - 8\pi J,
\yy                                                    \label{eq-F1}
         \nabla_i \left (\e^{-2\gamma}P \phi^i\right)
	 		 	\eql 4\pi \rho_e \e^{-3\gamma},
\ear
  where $\nabla_i$ and the Laplace operator $\Delta = \nabla_i \nabla^i$
  are defined in terms of the metric $h_{ik}$.

  The nonzero components of the Ricci and Einstein tensors are
\bear
	R^0_0 \eql - \e^{2\gamma} \Delta \gamma,         \label{Ricci}
\nn
	R^k_i \eql - \e^{2\gamma}(^h R^k_i + 2\gamma_i \gamma^i
			- \delta^k_i \Delta \gamma);
\yy                                                      \label{Einst}
	G^0_0 \eql \e^{2\gamma} (\half ^h R + \gamma^i \gamma_i
			- 2 \Delta \gamma),
\nn
	G_i^k \eql \e^{2\gamma} (- ^h R_i^k + \half ^h R \delta_i^k
			-2 \gamma_i \gamma^k +\delta_i^k \gamma_j \gamma^j),
\ear
  where $^h R^k_i$ and $^h R$ are the Ricci tensor and scalar obtained from
  the metric $h_{ik}$.

  The nonzero components of the SETs are
\bear                                                       \label{SET-s1}
	8\pi T^0_0{}_{(s)} \eql \eps \e^{2\gamma} \chi_i \chi^i,
\nn
	8\pi T_i^k{}_{(s)} \eql \eps \e^{2\gamma}
			(-2 \chi_i \chi^k + \delta_i^k \chi_j \chi^j);
\\
	8\pi T^0_0{}_{(e)} \eql P(\chi)\phi_i \phi^i,        \label{SET-F1}
\nn
	8\pi T_i^k{}_{(e)} \eql P(\chi)
			(2 \phi_i \phi^k - \delta_i^k \phi_j \phi^j);
\\                                                	    \label{SET-m1}
	T\mN{}_{(m)} \eql \rho_m \diag(1,\ 0,\ 0,\ 0).
\ear

  Let us now assume that the spatial metric $h_{ik}$ is flat (but not
  necessary written in Cartesian coordinates in which $h_{ik}=\delta_{ik}$).
  Then $^h R_i^k =0$, and the Einstein equations $R^0_0 = \ldots$ and $R_i^k
  = \ldots$ read
\bear
     	\e^{2\gamma} \Delta\gamma                            \label{ER00}
	    	\eql P \phi_i \phi^i + 4\pi \rho_m,
\yy                                                          \label{ERik}
       \e^{2\gamma} [2\gamma_i \gamma^k - \delta_i^k \Delta\gamma]
       \eql - 4\pi \rho_m \delta_i^k
       		+ P [2\phi_i\phi^k - \delta_i^k \phi_j\phi^j]
       			- 2\eps \e^{2\gamma} \chi_i \chi^k.
\ear
  Excluding $\rho_m$ from (\ref{ERik}), we obtain the first-order equation
\beq                                                          \label{int}
	\e^{2\gamma} (\gamma_i \gamma^k + \eps \chi_i \chi^k)
			= P\, \phi_i \phi^k
\eeq
  which does not contain the densities, hence it holds both in vacuum and in
  matter.

  Since in our static system $\rho_m u^\alpha \nabla_\alpha u_\mu$ reduces to
  $ -\rho_m \gamma_i$ (with the index $\mu = i$), the equation of motion
  (\ref{motion}) now leads to the equilibrium condition
\beq                                                          \label{equi}
	  \rho_m \gamma_i - \rho_e \phi_i \e^{-\gamma} = J\chi_i;
\eeq
  the continuity equation takes the usual form $\nabla_\mu(\rho_m u^\mu)=0$
  and holds automatically.

\section {The Majumdar-Papapetrou system and its scalar counterpart}

  Let us briefly recall the classical case $\chi \equiv 0$, $P \equiv 1$,
  that is, electrically charged dust, the Majumdar-Papapetrou (MP) system.
  The scalar equation (\ref{eq-s1}) is absent, while \eq (\ref{int}) gives
\beq                                                      \label{int0}
	\e^{2\gamma}\, \gamma_i \gamma^k = \phi_i \phi^k.
\eeq
  In particular, we can write $\e^{2\gamma} \gamma_1^2 = \phi_1^2$ and
  similar relations for the derivatives in $x^2$ and $x^3$, whose integration
  leads to
\bear
	 \e^{\gamma} \eql \pm \phi + C_1(x^2, x^3)
\nn
	 	     \eql \pm \phi + C_2(x^1, x^3)
\nn
	 	     \eql \pm \phi + C_3(x^1, x^2)
\ear
  whence it follows $C_1 = C_2 = C_3 = \const = 1$ according to the boundary
  conditions that $\phi \to 0$ and $\e^\gamma \to 1$ at spatial infinity.
  Thus we have
\beq                                                      \label{phi-MP}
	\pm \phi = \e^\gamma -1.
\eeq
  Using \eqs (\ref{eq-F1}) and (\ref{ER00}) as expressions for
  $\rho_e$ and $\rho_m$, we then obtain $\rho_\e = \pm \rho_m$ and
\beq
	4\pi \rho_m = \pm 4\pi \rho_e                     \label{rho-MP}
			= -\e^{3\gamma}\Delta(\e^{-\gamma}).
\eeq
  \eqs (\ref{int}) and (\ref{equi}) hold automatically. The charge and mass
  densities, equal in absolute value, are thus (up to the factor
  $\e^{3\gamma}$) sources of the effective gravitational potential
  $\e^{-\gamma}$. On the contrary, for each smooth $\gamma=\gamma(x^i)$ one
  finds the proper charge and mass distribution able to be its source.

  Noteworthy, it is not necessary here to assume a relationship between the
  functions $\phi(x^i)$ and $\gamma(x^i)$, it has emerged automatically from
  the equilibrium conditions.

  A somewhat similar situation is observed if we put, on the contrary,
  $\phi = \const \then F\mn \equiv 0$. In this case, \eq (\ref{int}) reads
  $\gamma_i\gamma^k + \eps \chi_i \chi^k =0$, whence it follows $\eps =-1$
  (an equilibrium is only possible with a phantom scalar field, which
  confirms its repulsive nature) and
\beq
	\pm \chi_i = \gamma_i.                             \label{chi-SMP}
\eeq

  Then \eqs (\ref{ER00}) and (\ref{equi}) express the densities $\rho_m$ and
  $J$ in terms of $\gamma(x^i)$:
\beq                                                     \label{rho-SMP}
	\pm 4\pi J = 4\pi \rho_m = \e^{2\gamma} \Delta \gamma.
\eeq
  As in the MP system, for each smooth $\gamma=\gamma(x^i)$ one finds the
  proper scalar charge and mass density distributions able to be its source.

  In both cases (\ref{rho-MP}) and (\ref{rho-SMP}) one can consider a set of
  point sources instead of their continuous distribution. However, the
  nature of these point sources is quite different. Thus, in the
  MP system the field of a point source satisfies $\Delta\e^{-\gamma} =0$,
  whence
\beq                                                     \label{gam-RN}
	\e^{\gamma} = x/(m + x), \cm x := (x_i x^i)^{1/2},
\eeq
  under the boundary condition $\gamma \to 0$ as $x \to \infty$, and $m$ has
  the meaning of the Schwarzschild mass. As is well known, it is an extreme
  \RN\ \bh\ with a horizon at $x=0$, at which the spherical radius
  $r = x\e^{-\gamma}$ is equal to $m$.

  Unlike that, in the Einstein-scalar field system \eq (\ref{rho-SMP})
  implies $\Delta\gamma =0$ outside a source, hence, under the same boundary
  condition, $\gamma = - m/x$, and the source at $x = 0$ is a naked
  singularity.

\section {General static configurations}

  In the general static case we have the following independent equations:
  (\ref{eq-s1}), (\ref{eq-F1}), (\ref{ER00}) and (\ref{int}). The tensor
  equation (\ref{int}) implies that $\gamma$, $\chi$ and $\phi$ are
  functionally related, so that, if $\gamma \ne \const$, we can put
  $\phi = \phi(\gamma)$ and $\chi = \chi(\gamma)$.

  Indeed, if we assume the contrary, e.g., that $\gamma$ and $\chi$ are
  functionally independent, then they can be chosen as curvilinear
  coordinates in 3-space, say, $\gamma = x^1$ and $\chi = x^2$. Then from
  the corresponding components of (\ref{int}) we immediately obtain
  $\phi_3^2 =0$ and $\phi_1 \phi_2 =0$, so that either $\phi=\phi(\gamma)$ or
  $\phi = \phi(\chi)$. If we take $\phi = \phi(\gamma)$, the $({}_{22})$
  component of \eq (\ref{int}) makes a contradiction (the l.h.s. is nonzero
  while the r.h.s. is zero), while if $\phi = \phi(\chi)$, the same happens
  to the $({}_{11})$ component. Thus our assumption, that $\gamma$ and
  $\chi$ are independent, is wrong. The same is concluded if we assume
  independence of $(\gamma, \phi)$ or $(\phi, \chi)$.

  Now, assuming $\phi = \phi(\gamma)$ and $\chi = \chi(\gamma)$, \eq
  (\ref{int}) takes the form
\beq                                                        \label{int00}
	\e^{2\gamma} (1 + \eps \chi_\gamma^2) = P \phi_\gamma^2,
\eeq
  where the index $\gamma$ denotes $d/d\gamma$.

  Hence we have the following arbitrariness: for any $P(\chi)$ and any
  3D profile $\gamma(x^i)$, even more than that, for an arbitrary scalar
  field distribution $\chi = \chi(\gamma)$ (up to certain restrictions),
  we find $\phi(\gamma)$ from (\ref{int00}), and the remaining field
  equations (\ref{eq-s1}), (\ref{eq-F1}) and (\ref{ER00}) give us the mass,
  electric and scalar charge distributions that support this field
  configuration. The condition (\ref{equi}) holds due to the Einstein
  equations.

  It is of interest whether or not it is possible to provide $J\equiv 0$,
  that is, to find configurations without an independent scalar charge,
  where the $\chi$ field exists only due to its interaction with the
  electromagnetic field.

  Assuming $J \equiv 0$, the equilibrium condition (\ref{equi}) gives
\beq
	\e^\gamma \rho_m = \rho_e\, d\phi/d\gamma.
\eeq
  Substituting it into (\ref{eq-s1}), we obtain the following relation:
\beq                                                           \label{J=0}
	2\eps \chi_\gamma\Delta \gamma + \biggl(
	2\eps \chi_{\gamma\gamma} + P_\chi\frac{\rho_m^2}{\rho_e^2}\biggr)
		\gamma_i \gamma^i = 0.
\eeq
  If we wish to admit configurations with any $\gamma(x^i)$, then the
  quantities $\Delta\gamma$ and $\gamma_i \gamma^i$ are actually
  independent, and we should require that the factors near them vanish.
  Then we obtain $\chi = \const$ (the scalar field is trivial) and
  $P = \const$ (no scalar-electromagnetic interaction), thus actually
  returning to the MP situation.

  We conclude that sufficiently general configurations with $P = P(\chi)$
  cannot be obtained without a scalar charge density $J$; however, special
  solutions to \eq (\ref{J=0}) with particular profiles of $\gamma$ for
  specific $P(\chi)$ must exist.

  Meanwhile, static dust with non-interacting $\chi$ and $F\mn$,
  corresponding to the special case $P \equiv 1$, can evidently form
  nontrivial configurations if we admit $J \ne 0$, and some examples will be
  given below.

  In what follows we will try to obtain examples of BH and quasi-BH (QBH)
  configurations in the simplest cases of spherical and \cy\ symmetries, and
  of special interest can be those of them where all kinds of matter are
  ``normal'', in particular, $\eps = +1$ and $\rho_m \geq 0$. Such an
  opportunity is not forbidden, but its realization causes some difficulty.
  The latter is illustrated by the expression for $\rho_m$ in terms of
  $\gamma(x)$ and $\chi(x)$, obtained by combining (\ref{ER00}) and
  (\ref{int}):
\beq                                                           \label{rho_m}
	4\pi \rho_m = \e^{2\gamma} (\Delta\gamma - \gamma^i\gamma_i
		-\eps\chi^i\chi_i).
\eeq
  We see that a normal scalar field ($\eps = 1$) makes a negative
  contribution to $\rho_m$. If, for example, we choose $\gamma(x^i)$
  satisfying the MP vacuum equation $\Delta (\e^{-\gamma}) =0$, we obtain
  simply $\rho_m \propto -\eps$.

  Nevertheless, since a normal scalar field, like gravity, creates
  attractive forces between dust particles, it is natural to expect the
  existence of phantom-free DMP systems where this new attractive component
  is balanced by an electric charge density larger than in MP systems.

\section{Spherically symmetric matter distributions: General features}

\subsection {Field equations and asymptotic behavior}

  In the case of spherical symmetry, the metric (\ref{ds}) with the flat
  3-metric $h_{ik}$ takes the form
\beq                                                          \label{ds1}
	ds^2 = \e^{2\gamma} dt^2 - \e^{-2\gamma}(dx^2 + x^2 d\Omega^2),
\eeq
  where $x$ is the usual spherical coordinate in flat 3-space and
  $d\Omega^2$ is the line element on a unit sphere. Our set of equations
  takes the form
\bear                                                         \label{chi1}
	2\eps x^{-2}\e^{2\gamma}(x^2\chi')'
			+ P_\chi \phi'{}^2 \eql - 8\pi J(x),
\yy                                                           \label{phi1}
	x^{-2} \big(P \e^{-2\gamma} x^2 \phi'\big)'
			\eql 4\pi \rho_e \e^{-3\gamma},
\yy                                                           \label{gam1}
	x^{-2} \e^{2\gamma} (x^2\gamma')' - P \phi'{}^2
			\eql 4\pi \rho_m,
\yy                                                           \label{int1}
	\gamma'{}^2 + \eps \chi'{}^2 \eql \e^{-2\gamma} P \phi'{}^2,
\yy                                                           \label{equi1}
	\rho_m \gamma' - \rho_e \phi' \e^{-\gamma} \eql J\chi',
\ear
  where the prime denotes $d/dx$. The above arbitrariness transforms here
  into the freedom of choosing the functions $\gamma(x)$ and $\chi(x)$
  even if the coupling function $P(\chi)$ has been prescribed from the
  outset. All other quantities are then found from \eqs
  (\ref{chi1})--(\ref{equi1}), including the mass, electric and scalar
  charge distributions that support this field configuration.

  It is of interest how to choose the arbitrary functions in order to obtain
  a starlike configuration with a regular center or a BH. In all cases it is
  also of interest to see how to provide a phantom-free configuration with
  $\eps = +1$ and $\rho_m \geq 0$. In doing so, we can use \eq (\ref{rho_m})
  which now takes the form
\beq                                                          \label{rho_m1}
	4\pi \rho_m = \e^{2\gamma} (\gamma'' + 2\gamma'/x
			- \gamma'{}^2 -\eps\chi'{}^2).
\eeq

\noi
  {\bf A regular center} is obtained in the metric (\ref{ds1}) at $x=0$
  if and only if
\beq                                                         \label{gam-c}
  	\gamma(x) = \gamma_c + O(x^2), \cm \gamma_c = \const,
\eeq
  and we can use the Taylor expansion
\beq                                                          \label{A-c}
     \e^{2\gamma} \equiv A(x) = A_0 + \half A_2 x^2 + \fract{1}{6} A_3 x^3
	+ \cdots,
	\ \ \ A_i = \const, \ \ \ A_0 = \e^{2\gamma_c}.
\eeq
  According to the field equations, we must also have, as $x \to 0$,
\beq                                                         \label{phi-c}
	\phi(x) = \phi_c + O(x^2), \cm \chi(x) = \chi_c + O(x^2),
\eeq
  with some constants $\phi_c$ and $\chi_c$, and this provides finite values
  of the densities $\rho_m$, $\rho_e$ and $J$ at the center. The expression
  (\ref{rho_m1}) then leads to
\beq                                                         \label{rho_mc}
	8\pi \rho_m = 3A_2 + 2 A_3 x + \cdots,
\eeq
  while the $\chi$ field only contributes beginning with $O(x^2)$. Thus a
  positive density near the center requires, as usual, that $g_{00} = A(x)$
  should have there a minimum.

\medskip\noi
{\bf Horizons.}
  The question on possible Killing horizons inside a charged dust
  distribution is somewhat more involved. To begin with, in (\ref{ds1}) the
  coordinate $x$ is quasiglobal, which means that it behaves near a
  horizon in the same way as null Kruskal-like coordinates \cite{BR-book},
  whence follows the requirement that all functions characterizing the
  system should be analytical at the corresponding point $x = \xh$, hence
  near the horizon $\e^{2\gamma}\approx B_0 (x-\xh)^n$, where
  $B_0 = \const > 0$ and $n \in \N$ is the order of the horizon.

  From (\ref{ds1}) it is clear that a horizon of finite radius
  $\rh = x\e^{-\gamma}\big|_{x=\xh}$ is only possible with $\xh = 0$ and
  $n =2$ (a {\it double, or extremal horizon}). Assuming a simple horizon
  ($n=1$) at $x =0$, we obtain $\rh =0$, hence a singularity; on the other
  hand, a higher-order horizon ($n > 2$) with any $\xh \geq 0$ corresponds
  to an infinite radius $\rh$. A continuation beyond such a horizon is
  possible under some special conditions, and then we are dealing with the
  so-called ``cold BHs'', see \cite{cold}.

  We will not consider here such an unusual opportunity and thus assume that
  near $x = 0$
\beq                                                          \label{A-hor}
	\e^{2\gamma} \equiv A(x) = \half A_2 x^2 + \fract{1}{6} A_3 x^3
			+ \fract{1}{24} A_4 x^4 + \cdots,
	\ \ \ A_i = \const, \ \ \ A_2 > 0.
\eeq

  Furthermore, the metric analyticity leads, via the Einstein equations, to
  analytic behaviors of the SET components, in particular, of the quantities
  $P \phi'{}^2$ and $\e^{2\gamma} \chi'{}^2$. The latter means that we can
  in principle assume
\beq
	\e^\chi \sim x^s, \ \ \ s \in \R \ \ \ {\rm as} \ \ x\to 0,
\eeq
  so that $\chi' \approx s/x$, and only $s=0$ corresponds to
  $|\chi'| < \infty$ as $x \to 0$. It follows from (\ref{int1}) that
\beq
	P \phi'{}^2 \approx \half A_2 \eps s^2\ \ \ {\rm as}\ \ x\to 0.
\eeq
  Then (\ref{gam1}) gives for the matter density
\beq                                                        \label{rho_mhs}
	2\pi \rho_m (0) = - A_2 \eps s^2.
\eeq

  Thus if $s \ne 0$, then $\rho_m < 0$ in the presence of a normal
  scalar field ($\eps=+1$) and positive with a phantom one ($\eps = -1$).

  The behavior of the other densities, $\rho_e$ and $J$, at the horizon
  depends on the choice of the coupling function $P(\chi)$.
  \eq (\ref{int1}) gives $P\phi'{}^2 \to \half A_2(1 + \eps s^2)$ as
  $x\to 0$, so that the normal, positive sign of the coupling function
  $P(\chi)$ requires $|s| < 1$ if $\eps = -1$. In the special case
  $\eps = -1$ and $s=1$, we obtain $P\phi'(0)^2 = 0$. Let us, however,
  discuss the generic case where $1 + \eps s^2 \ne 0$.

  Assuming that $P(\chi)$ is finite as $x\to 0$ (while $\e^\chi \sim x^s$),
  we see that $\phi'(0)$ is finite, then from (\ref{phi1}) it follows that
  generically $\rho_e \sim x$, but there can be special cases where $\rho_e
  = o(x)$ (if the derivative of the expression in parentheses is zero at
  $x=0$). From (\ref{equi1}) we find that the scalar charge density $J$ is
  finite at the horizon.

  In the important case $P(\chi) = \e^{2\lambda\chi}$, $\lambda = \const\ne
  0$ (or if $P(\chi)$ behaves like that near the horizon) we have $P \sim
  x^{2\lambda s}$ as $x\to 0$. From (\ref{int1}) we then find that $\phi'
  \sim x^{-\lambda s}$. \eq (\ref{phi1}) yields in turn $\rho_e \sim
  x^{\lambda s}$, and from (\ref{equi1}) it follows that $J$ tends to a
  finite limit.  The requirement $|\rho_e| < \infty$ implies $\lambda s >
  0$ (both $\lambda$ and $s$ are nonzero by assumption), hence $\rho_e$
  vanishes at the horizon as well as $P(\chi)$, while $\phi'$ is infinite.

  Let us now discuss the case $s=0$, such that $\chi'$ and $\chi$ are
  finite at the horizon. At small $x$ we then obtain $\rho_m \sim x^2$, and
  its sign does not directly correlate with $\eps$. A substitution of
  (\ref{A-hor}) into (\ref{rho_m1}) leads to
\beq                                                        \label{rho_mh}
	8\pi \rho_m = \biggl(-\frac{A_3^2}{12 A_2} + \frac{A_4}{12}
			- \eps \chi'(0)^2 \biggr)x^2 + O(x^3).
\eeq
  From (\ref{int1}) we find that $\phi'$ (and hence $\phi$) are finite at
  the horizon provided that $P(\chi)$ is finite (notably, now the assumption
  $P (\chi) = \e^{2\lambda\chi}$ belongs to this generic case). \eq
  (\ref{gam1}) then leads, as before, to $\rho_e \sim x$ or possibly $\rho_e
  = o(x)$, while $J$ generically tends there to a finite limit. We conclude
  that such configurations, which are in general perfectly regular and
  smooth, still contain an anomaly: the density ratios $\rho_e/\rho_m$ and
  $J/\rho_m$ are infinite at the horizon.

\medskip\noi
{\bf Asymptotic flatness.}
  Next issue is the asymptotic behavior of the system at large $x$. If we
  consider, as usual, finite dust distributions, the internal domain with
  nonzero densities should be matched to an external domain described by the
  corresponding ``vacuum'' EMD solution; however, such solutions to the
  field equations are only known for some special
  choices of $P(\chi)$, e.g., $P = \e^{2\lambda\chi}$.

  Therefore, instead of dust balls of finite size placed in vacuum, we will
  consider \asflat\ matter distributions with a smoothly decaying density.
  Thus we can take at large $x$
\beq                                                         \label{A-as}
       A(x) = 1 - \frac{2m}{x} + \frac{q_*^2}{x^2} + \cdots,
       \cm   \chi(x) = \chi_\infty + \frac{\chi_1}{x} + \cdots,
\eeq
  and \eq (\ref{rho_m1}) then yields
\beq                                                         \label{rho_ma}
       4\pi \rho_m = \frac{1}{x^4} (-3m^2 + q_*^2 - \eps \chi_1^2)
       			+ o(x^{-4}).
\eeq
  This expression clearly shows that large charges $q$ are necessary for
  obtaining $\rho_m > 0$ if $\eps = +1$. (Note that the extreme \RN\
  solution (\ref{gam-RN}) with the charge $q =m$ corresponds in the notation
  (\ref{A-as}) to $q_*^2 = 3m^2$.)

  The densities $\rho_e$ and $J$ also behave in general as $1/x^4$ at
  large $x$.

\medskip\noi
{\bf Integral charges.} The field at flat spatial infinity is characterized
  by integral charges: the electric charge $q$ such that the electric field
  strength is $\phi' = q/x^2 + o(1/x^2)$, the scalar charge $D$ such that
  $\chi' = D/x^2 + o(1/x^2)$, and the mass $m$ corresponding to the
  Schwarzschild asymptotic $\e^\gamma \approx 1-m/x$, hence $\gamma'
  \approx m/x^2$ (note that $x \approx r$ at large $x$).  A relation between
  these three quantities directly follows from \eq (\ref{int1}). Indeed,
  multiply (\ref{int1}) by $x^4$ and take the limit $x \to \infty$ to obtain
\beq
	 m^2 - q^2 + \eps D^2 =0,                      \label{charges}
\eeq
  where we have taken into account that both $\e^\gamma$ and $P$ tend to
  unity (the latter one due to the requirement that a weak electromagnetic
  field should be Maxwell). This generalizes a similar relation (2.12) from
  \cite{clem2}, written there for vacuum EMD systems with an exponential
  coupling function $P(\chi)$. (Unlike (\ref{charges}), in \cite{clem2} there
  stands $\eta_2 = \sign P$ before $q^2$; we obtain the same if we also
  admit an anti-Maxwell asymptotic behavior, $P\to -1$, of the
  electromagnetic field.)

  Thus, as compared to the MP system where $q = \pm m$ (not only the
  densities, but also the integral mass and charge are equal in absolute
  value), a balance in the DMP system requires $m^2 > q^2$ in the presence
  of a phantom scalar with $\eps = -1$ (both electric and scalar fields are
  repulsive), but $m^2 < q^2$ in the presence of a canonical scalar field
  which is attractive like gravity.

  Evidently, the relation (\ref{charges}), having been derived from the
  asymptotic properties of \sph\ systems, is valid as well for any \asflat\
  (island-like) EMD system since all of them are approximately \sph\ in the
  asymptotic region.

\subsection{Some relations in Schwarzschild coordinates}

  In the present paper, we mainly consider \sph\ systems with the metric
  (35) written using the coordinate $x$ which is, by the existing
  terminology, simultaneously isotropic (since the spatial part is
  conformally flat) and quasiglobal (since $g_{tt} g_{xx} = -1$
  \cite{BR-book}). However, bearing in mind the popularity of the
  Schwarzschild (curvature) coordinates, including their usage in previous
  papers on MP systems and QBHs, we here present some general relations in
  these coordinates, which can be helpful for comparison.

  The Schwarzschild coordinate $r$ is the curvature radius of coordinate
  spheres, i.e.,
\beq                                                          \label{r}
	r = x \e^{-\gamma}\ \ \then \ \ \e^{\gamma} = x/r,
\eeq
  so that in $e^{\gamma}$ the factor $x$ responsible for the possible
  existence of a horizon (at $x=0$) is singled out. The metric now reads
\beq
	ds^2=\e^{2\gamma}dt^2 - \e^{-2\gamma}\psi^2 dr^2 - r^2 d\Omega^2,
\eeq
  where
\beq
	\psi = \frac {dx}{dr} = \frac{d(re^{\gamma})}{dr}.
\eeq
  For extremal BHs considered here, $e^{\gamma}\sim r-h$ near the horizon
  $r = h$, hence $\psi$ is finite there.

  A straightforward calculation using (\ref{rho_m1}) gives for the matter
  density
\beq						\label{rho_mr}
	4\pi\rho_m = \frac{\e^{2\gamma}}{\psi^2}
	\biggl[\frac{\psi_r}{\psi r} - \eps \chi_r^2\biggr],
\eeq
  where the subscript $r$ means $d/dr$.

  Let the function $x(r)$ be monotonically increasing, then $\psi >0$. In the
  case of a normal dilaton field ($\eps = +1$), a necessary condition for a
  positive matter density is $\psi_r > 0$, and since the first term
  in the brackets in (\ref{rho_mr}) is finite at $r=h$, the quantity
  $\chi_r$ is also finite. Then from (\ref{rho_mr}) it follows, in agreement
  with (\ref{rho_mh}).
\beq
	4\pi\rho_m = O\big((r-h)^2\big)
\eeq
  This argument does not work for phantom fields ($\eps =-1$): in that case
  $\chi_r$ is not necessarily finite at the horizon, hence $\rho_m$ may behave
  as $O(r-h)$ or be finite there.

  In the absence of a horizon, the condition $e^{\gamma}=1+O(r^2)$ at small
  $r$ ensures the existence of a regular center.

\subsection {Quasi-black holes}

  In MP systems without a dilaton field, to describe a configuration
  completely, it is sufficient to specify the metric function $\gamma(x^{i})$.
  All other quantities are then found unambiguously. Under some natural
  conditions, such as a nonnegative matter density, any $\gamma(x^{i})$ is
  suitable. As a result, one can choose this function in such a way that the
  system belongs to an interesting class of static (thus non-collapsing)
  configurations called quasi-black holes (QBHs). In the space of parameters,
  such systems are arbitrarily close to the threshold of forming a horizon,
  but nonetheless a horizon is absent. A compact body of this kind does not
  collapse even for an arbitrary small difference between the boundary and
  the gravitational radius.  (A typical example are so-called Bonnor stars
  made of charged dust with $|\rho_e| = \rho_m$ \cite{bon99}.) An important
  feature of QBHs is that they have arbitrarily large redshifts for signals
  emitted from a certain region, as if it were a neighborhood of a BH
  horizon. The whole system then looks completely like a BH for a distant
  observer, although it is quite different from a BH in its total structure
  \cite{qbh}.

  For a discussion of the main properties of QBHs one can consult \cite{qbh}.
  Below, we will see that such objects can also exist in DMP systems, and we
  will discuss their basic features.

  By definition, it is assumed that in some region $r\leq r^*(c)$ of a QBH
  $\e^{\gamma} \sim c$, where $c$ is a small parameter, and it is also
  usually assumed (though it is not quite necessary) that $r^*(0)$ is the
  horizon radius of a black hole that corresponds to the limit $c\to 0$.
  Thus the most general \ssph\ QBH in our problem setting is a system with
  the metric (\ref{ds1}) and a regular center, such that the Taylor
  expansion (\ref{A-c}) at small $x$ is specified in terms of $c$ as
\beq
     \e^{2\gamma} \equiv A(x,c) = A_0 (c) + \half A_2 (c) x^2 + \ldots,
\eeq
  where $A_0 (c) \to 0$ as $c \to 0$ while $A_2(0)$ is finite. One can
  re-denote the parameter $c$ in such a way that $A_0 (c) = \half A_2(0)
  c^2$, and as a result, without loss of generality, one can assume
\beq
	\e^{2\gamma} = \frac{x^2 + c^2}{f^2(x,c)},        \label{gam-q}
\eeq
  where $f$ is a smooth function that has a well-defined and nonzero limit
  $c\to 0$. (One can note that this representation is quite equivalent
  to the one used in some previous papers on QBHs \cite{lemos2,lemos3},
  viz., $e^{2\gamma} = c^2 F(x)$. However, if the parameter $c$ enters
  into the function $F$, that is, $F = F(x,c)$, the situation can be
  more subtle.)

  As already mentioned, if we simply put $c=0$ in (\ref{gam-q}), we obtain
  a BH metric with a horizon at $x=0$. In particular, taking $f(x,0) = x+m$,
  we obtain the extreme \RN\ metric. In the special case
  $f(x,c) = q +\sqrt {x^2+c^2}$, we have the example given by \eq (17) of
  \cite{qbh}. At small enough $c$ and $x\lesssim c$, $e^{2\gamma} = O(c^{2})$
  is arbitrarily small.

  It is of interest that, with the metric ansatz (\ref{gam-q}), the region
  where the ``redshift function'' $\e^\gamma$ is small, is itself not small
  at all. Indeed, let us choose such a unit of length that $f(x,c) = O(1)$
  and $c \ll 1$. Consider the sphere $x=c$, which certainly belongs to the
  high redshift region since $\e^{\gamma}\big|_{x=c} = \sqrt{2}c/f(c,c) \ll
  1$. Then the radius $r(c)$ of the same sphere is $f(c,c)/\sqrt{2} =
  O(1)$; the distance from the center to this sphere is
\beq
	l(0,c) = \int_0^c \e^{-\gamma}\,dx
		= \overline{f}(c) \ln(1+\sqrt{2}) = O(1),
\eeq
  where $\overline{f}(c)$ is a certain mean value of the function $f(x,c)$
  on the interval $(0,c)$. Lastly, the volume of the ball $x \leq c$ is
\beq
	V(c) = 4\pi \int_0^c \e^{-3\gamma}x^2\,dx
	     = 4\pi \overline{f^3}(c)[\ln (1+\sqrt{2}) - 1/\sqrt{2}]
	     \approx 2.19\ \overline{f^3}(c),
\eeq
  where $\overline{f^3}(c)$ is a certain mean value of $f^3(x,c)$ on the
  interval $(0,c)$.

\subsection {Limiting transitions for quasi-black holes}

  A straightforward transition $c \to 0$ in (\ref{gam-q}) gives
  $\e^{\gamma} = x/f(x,0)$, which leads to an extreme BH metric according to
  the above description since we have assumed that $f(x.c) = O(1)$ at small
  $c$. Evidently, if the QBH metric with this $f(x.c)$ is \asflat, then the
  resulting BH metric will also be \asflat.

  There are, however, some interesting variants of the transition
  $c \to 0$ which can be constructed by analogy with \cite{qbh} and
  correspond to imposing certain additional requirements on the resulting
  space-time properties. This can be achieved by introducing special
  $c$ dependences for the coordinates and parameters of our QBH metric.
  Mathematically, we are dealing with the so-called limits of space-time
  procedure \cite{ger}.

{\bf First,} let us try to preserve a regular center while $c\to 0$.
  To do so, we introduce the new coordinates
\beq
	X = x/c, \cm T = ct/f_0, \cm f_0 := f(0,0),    		\label{sub1}
\eeq
  and, having rewritten the metric (\ref{ds1}) with (\ref{gam-q}) in terms
  of these new coordinates, consider $c \to 0$. The resulting metric reads
\beq                                                        \label{ds-lim1}
	ds^2 = (1 + X^2)dT^2 - \frac{f_0^2}{1+X^2} (dX^2 + X^2 d\Omega^2).
\eeq
  This limiting metric is geodesically complete, it still has a regular
  center at $X =0$, but it is not \asflat: instead, at large $X$ it approaches
  a flux-tube metric with $r(X) = \const$ and $g_{TT} \to 0$, as if there
  were a horizon infinitely far away. The metric (\ref{ds-lim1}) may be
  interpreted as a result of infinitely stretching the neighborhood of
  the center. Noteworthy, it does not depend on the particular form of the
  function $f(x,c)$ and even on $f(x,0)$: only the constant $f_0$ remains.

  Assuming that the functions $f(x,c)\ne 0$ and $\chi'(x)$ are finite and
  smooth at $x=0$, a substitution of the ansatz (\ref{gam-q}) and
  (\ref{sub1}) into the field equations shows that the matter density tends
  to a finite limit as $c \to 0$, namely,
\beq
	4\pi \rho_m \to \frac{3}{f_0^2 (1 + X^2)};
\eeq
  the quantities $\phi'$, $\rho_e$ and $J$ also remain finite. (Note that
  while $\chi' \equiv d\chi/dx$ is finite, $d\chi/dX = c \chi' \to 0$.)

{\bf Second}, following \cite{qbh}, we can introduce the notion of a
  quasi-horizon as a minimum of the function $g^{rr}$, where $r$ is, as
  before, the Schwarzschild coordinate: this minimum is zero in the case of
  an extreme BH and has a small positive value in a QBH. It can be
  verified that with (\ref{gam-q}) the quasi-horizon corresponds to
  $x = x^* \sim c^{2/3}$ while the corresponding value of $r$ approximately
  coincides with the genuine BH horizon radius obtained in the
  straightforward transition $c \to 0$.

  We can try to construct such a transition $c\to 0$ that will somehow
  preserve the geometry near the quasi-horizon. In accord with the
  above-said, we substitute instead of $x$ and $t$
\beq
       \xi = c^{-2/3} x, \cm \tau = f_0^{-1} c^{2/3} t,     \label{sub2}
\eeq
  and consider the limit $c\to 0$ in these new coordinates. The result is
\beq							\label{ds-lim2}
       ds^2 = \xi^2 d\tau^2 - \frac{f_0^2}{\xi^2} d\xi^2 -f_0^2 d\Omega^2.
\eeq
  It is a pure flux-tube metric, coinciding up to a constant factor with
  the Bertotti-Robinson metric \cite{br1,br2} in agreement with \eq (30)
  of \cite{qbh}; it can be called an infinitely long throat.
  Its interesting feature is that, although $r$ is constant, the metric is
  strongly $\xi$-dependent. Instead of a regular center, it contains a
  second-order horizon at $\xi = 0$, and the region $\xi < 0$ beyond it is
  an exact copy of the region $\xi > 0$. Let us recall that the
  Bertotti-Robinson metric approximately describes the throat-like
  neighborhood of the extremal RN metric, but, at the same time, it is an
  exact solution to the Einstein-Maxwell equations and can be considered
  without reference to the RN metric.

  It can be directly verified that (again assuming finiteness and
  smoothness of $f(x,c)\ne 0$ and $\chi'(x)$), with the substitution
  (\ref{sub2}) the limiting values of all densities $\rho_m,\ \rho_e,\ J$ as
  $c \to 0$ are zero; however, there is a finite limiting value of $\phi'$
  [namely, $\phi'{}^2 \to 1/(P_0 f_0^2)$, where $P_0 = P(\chi(0))$]. This
  looks natural since the Bertotti-Robinson metric is known to be supported
  by a pure Maxwell field.

  In fact, the metric (\ref{ds-lim2}), coinciding with the near-horizon
  approximation of the extremal RN metric, is also similar to the same in
  scalar-tensor cold black holes \cite{cold,cold-q}. This near-horizon
  geometry seems to be quite general for black holes with zero Hawking
  temperature. Its two-dimensional part (leaving aside the angular part)
  has the same structure as the corresponding AdS metric. Its curious aspect
  is that the distance from any point to the horizon is infinite. In
  \cite{an1,an2} it has been speculated that this property may be a reason
  for an anomalous behavior of quantum fields near extremal black holes.

  We conclude that the straightforward limit $c \to 0$ concerns the whole
  QBH space-time but well preserves its properties only outside the
  quasi-horizon, including asymptotic flatness. The other two transitions
  correspond to ``looking through a microscope'' at (a) the inner region
  corresponding to $0\leq x\lesssim c$, $0\leq r\lesssim c$, and (b) the
  intermediate region (near the quasi-horizon), $x = O(c^{2/3})$,
  $r-r^{\ast} = O(c^{2/3})$. It is possible to make a more detailed division
  (say, outside region (b)), but it is not necessary since at small but
  nonzero $c$ these three regions overlap and cover the whole space. It is
  of interest that the restricted regions (a) and (b) turn into geodesically
  complete space-times in the limit $c=0$.  The resulting metrics can be
  considered on their own rights as solutions to the field equations.

  The general properties of the metric with $\gamma(x)$ given by
  (\ref{gam-q}) in the limit $c\to 0$ are independent of the choice of the
  smooth finite function $f(x,c)$, quite similarly to what was described in
  Sec.\,III of \cite{qbh}. Since (\ref{gam-q}) is a general representation
  of $\e^{2\gamma}$ for QBH metrics, the limiting metrics (\ref{ds-lim1})
  and (\ref{ds-lim2}) are also universal for the whole present class
  of QBHs.

\section {Spherically symmetric matter distributions: Examples}

\subsection{Black holes}

\noi{\bf Example BH1.}
  The simplest extremal BH metric is obtained with the \RN\ ansatz
\beq
       \e^\gamma = \frac{x}{x+m},                           \label{RN}
\eeq
  where $m$ is the Schwarzschild mass equal to the total electric charge
  $q$. The familiar form of the extreme \RN\ metric is obtained from
  (\ref{ds1}) with (\ref{RN}) by substituting $x = r - m$, leading to
  $\e^{2\gamma} = (1 - r/m)^2$. The \RN\ metric is well known as an
  electrovacuum solution to the Einstein equations, corresponding, in terms
  of the present paper, to $\chi =0$, $P(0) = 1$, and all densities equal to
  zero. We shall see, however, that the same metric can be associated with
  dilatonic charged dust distributions.

  With (\ref{RN}), \eq (\ref{rho_m1}) gives simply
\beq                                                        \label{rho_m2}
	4\pi \rho_m = -\eps \e^{2\gamma} \chi'{}^2,
\eeq
  so that a positive mass density requires a phantom scalar $\chi$. A
  phantom-free configuration is thus impossible.

  Let us choose $\chi'(x)$ proportional to $\gamma'$, namely,
\beq                                                         \label{chi'2}
	\chi' = \frac{D}{x(x+m)}, \cm D=\const.
\eeq
  Then \eq (\ref{int1}) leads to
\beq                                                         \label{phi'2}
	P \phi'{}^2 = \frac{m^2 + \eps D^2}{(x+m)^4} \equiv
						\frac{C^2}{(x+m)^4},
\eeq
  where we assume $C \geq 0$, so that $P\phi'{}^2 \geq 0$. We then obtain
  the following expressions for the densities:
\bearr                                                       \label{dens2}
	4\pi \rho_m = -\frac{\eps D^2}{(x+m)^4},
		\cm
	4\pi \rho_e = \pm \frac{C x}{(x+m)^3} \frac{P'}{2\sqrt{P}},
\nnn
	4\pi J = - \frac{\eps D m}{(x+m)^4}
			- \frac{C^2 x}{D(x+m)^3} \frac{P'}{2P},
\ear
  with small and large $x$ behaviors in agreement with the general
  observations of the previous section, provided that $P(\chi)$ is a smooth
  function at all relevant values of $\chi$: thus, all densities are finite
  as $x\to 0$ while at large $x$ all of them behave as $x^{-4}$. (Note that
  $P' = \chi' dP/d\chi$, and, in particular, $\chi' \sim x^{-1}$ at small
  $x$, and $\chi' \sim x^{-2}$ at large $x$.)

  This solution is valid for arbitrary coupling functions $P(\chi) > 0$,
  including the important special cases $P \equiv 1$ and
  $P = \e^{2\lambda\chi}$. The same will be true in our other examples.

  Two special cases are worth mentioning. First, the case $C=0$ (that is,
  $\eps = -1$ and $D^2 = m^2$) corresponds to a vanishing electric field and
  $\rho_e \equiv 0$, so that we obtain the situation described by \eqs
  (\ref{chi-SMP}) and (\ref{rho-SMP}). Thus a BH with the \RN\ metric can be
  supported by a scalarly charged dust distribution instead of an
  electromagnetic field. On the other hand, if we assume $D =0$, all
  densities vanish, and we return to the familiar (electrovacuum) \RN\ BH.

\medskip\noi
{\bf Example BH2.}
  In the previous example either the scalar field or dust were
  necessarily of phantom nature. Let us show that our problem setting makes
  it possible to find an \asflat\ BH without phantoms. We now choose
\beq
	\e^\gamma = \frac{x}{m+ 2x -y}, \cm                     \label{A3}
	\chi' = \frac{b}{y^2}, \cm y := \sqrt{x^2 + a^2},
\eeq
  where $a,\ b,\ m$ are constants, and $m$ is again the Schwarzschild mass.
  characterizing the gravitational field at large $x$.

  The choice of $\chi'$ satisfies the condition $|\chi'(0)| < \infty$, which
  is necessary for $\rho_m > 0$ according to the previous section.
  With $\eps = +1$, we obtain from (\ref{A3})
\beq                                                         \label{rho_m3}
	4\pi \rho_m = \frac{x^2[(a^2+b^2)y - b^2(2x+m)]}{y^4 (2x-y+m)^3}.
\eeq
  This expression is positive at all $x > 0$ in a certain region of the
  parameter space. Thus, putting $m=1$ (which fixes the units) and $a=0.5$
  (for example), we find that $\rho_m > 0$ for $0 < b < b_0 \approx 0.369$,
  see Fig.\,1.

\begin{figure}
\centering
\includegraphics[width=7.5cm]{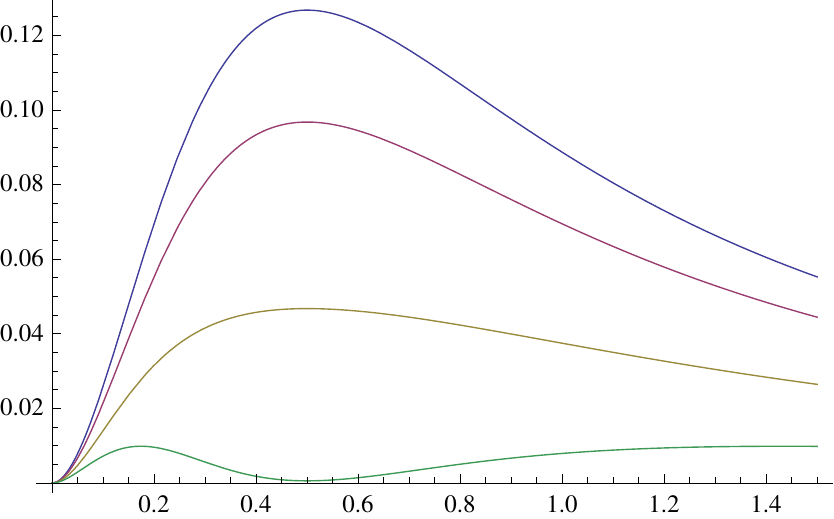}\ \
\caption{The density $\rho_m(x)$ in Example BH2 for $m =1$, $a = 0.5$, and
	$b = 0.1,\ 0.2,\ 0.3,\ 0.369$ (upside down).}
\end{figure}

  The expressions for $\rho_e$ and $J$ turn out to be very bulky and will
  not be reproduced here. It is only important that, for a generic choice of
  $P(\chi)$, they are everywhere finite and regular and behave at the
  horizon as described in the previous section.

\subsection{Quasi-black holes}

\medskip\noi
{\bf Example QBH1.}
  Let us modify the BH ansatz (\ref{RN}) by putting
\beq                                                          \label{RN-c}
     \e^{2\gamma} \equiv A(x) = \frac{z^2}{(m + z)^2},
     \cm 		z := \sqrt{x^2 +c^2},
\eeq
  where $m > 0,\ c > 0$ are constants, and $c=0$ returns us to the case
  (\ref{RN}). At large $x$ (\ref{RN-c}) leads to $A \approx 1 - 2m/x +
  3m^2/x^2 +...  $, hence $m$ has again the meaning of a Schwarzschild mass.
  Near the center $x = 0$ we have
\beq                                                         \label{q1-c}
      A(x) = \frac{c^2}{(m+c)^2} + \frac{mx^2}{(m+c)^3} + \ldots,
\eeq
  which confirms that the center is regular and also that with $m > 0$,
  according to (\ref{A-c}) and (\ref{rho_mc}), the matter density is
  positive at small $x$.

  Now, calculating the first three ($\gamma$-dependent) terms in
  (\ref{rho_m1}), we obtain
\beq                                                        \label{q1a}
	4\pi\rho_m = \frac{3m c^2}{z^2 (m+z)^3}
		       - \frac{z^2}{(m+z)^2} \eps \chi'{}^2.
\eeq

  According to (\ref{q1a}), no phantom scalar field is needed now to provide
  a positive matter density, and the simplest model can be built without
  scalars, in the Majumdar-Papapetrou framework. Indeed, consider the ansatz
  (\ref{RN-c}) along with $\chi'=0$ and $P \equiv 1$.  Then $\rho_m$ is
  given by the first term in (\ref{q1a}), \eq (\ref{phi1}) leads, as
  expected, to $\rho_e = \pm \rho_m$, and (\ref{equi1}) to $J \equiv 0$. The
  total charge is then equal to mass.

  To obtain a good example with the ansatz (\ref{RN-c}) but $\chi \ne
  \const$, let us choose $\chi'(x)$ in such a way as to provide regularity
  at $x=0$ (thus $\chi' \to 0$ as $x\to 0$) and positivity of matter density
  under the condition $\eps=+1$.  Since the first term in (\ref{q1a})
  behaves as $x^{-5}$ at large $x$, the same (or quicker) decay is required
  from the second term, that is, if $\chi' \sim x^{-n}$, then $n \geq 5/2$.
  For example, we can take
\beq
	\chi'(x) = b^2 x/z^4,
\eeq
  with a sufficiently small constant $b > 0$. It is easy to verify that
  $\rho_m > 0$ at all $x$ if we take $b \leq c < m$.
  A point of interest is that while in the BH case ($c=0$) only a phantom
  scalar $\chi$ can be compatible with $\rho_m > 0$, in the QBH case a
  canonical scalar ($\eps = 1$) is also admitted.

  The expressions for $\rho_e$ and $J$ for any $P(\chi)$ again turn out to
  be cumbersome and are not presented here.

\medskip\noi
{\bf Example QBH2.} Now we modify the expression (\ref{A3}) for $\gamma$ and
  put
\beq
	\e^\gamma = \frac{z}{m+ 2z -y},                        \label{A4}
	\cm y := \sqrt{x^2 + a^2},
	\cm z := \sqrt{x^2 + c^2},
\eeq
  with certain positive constants $m,\ a,\ c$. At small and large $x$ we
  have
\bear                                                         \label{A4_0}
	x\to 0: && \e^{2\gamma} = \frac{c^2}{(m-a+2c)^2}
			+ x^2\,\frac{m - a + c^2/a}{(m-a+2c)^3} + O(x^4),
\yy
	x\to \infty: && \e^{2\gamma} = 1 - \frac{2m}{x}       \label{A4_i}
			+ \frac{3m^2+a^2-c^2}{x^2} + O(x^{-3}).
\ear
  Such asymptotic behaviors mean that the configuration has a regular center
  and is \asflat; as before, $m$ has the meaning of the Schwarzschild mass.
  Moreover, assuming
\beq
	c < a < m,                                           \label{par4}
\eeq
  we can be sure that $\rho_m > 0$ near the center since $\e^\gamma$ has a
  minimum there (see Section 6). For $\rho_m$ we then obtain
\bearr                                                       \label{rho_m4}
      \rho_m = \frac{1}{y^3 z^2 (m - y + 2 z)^4}
			    	\Big[ 3 a^4 c^2 (-2m + y - 2z)
\nnnv \cm
	+ c^2 x^2 [3 m^2 y + x^2 (y -2z) + 2c^2 (m -y +2z) + m (-4x^2 + 6yz)]
\nnn \cm
	+ a^2 \{(x^4 + 3c^4)(m - y + 2 z) + c^2 [3m^2 y + 2x^2 (y -2z)
		+ m (-8x^2 + 6yz)]\} \Big]
\nnn \cm
	+ \frac{\eps z^2}{(m + 2z -y)^2} \chi'{}^2.
\ear
  Numerical calculations show that $\rho_m > 0$ for proper choices of the
  dilaton field profile $\chi(x)$ with $\eps=+1$ under the same condition
  (\ref{par4}) for the model parameters.
\begin{figure}
\centering
\includegraphics[width=6.2cm]{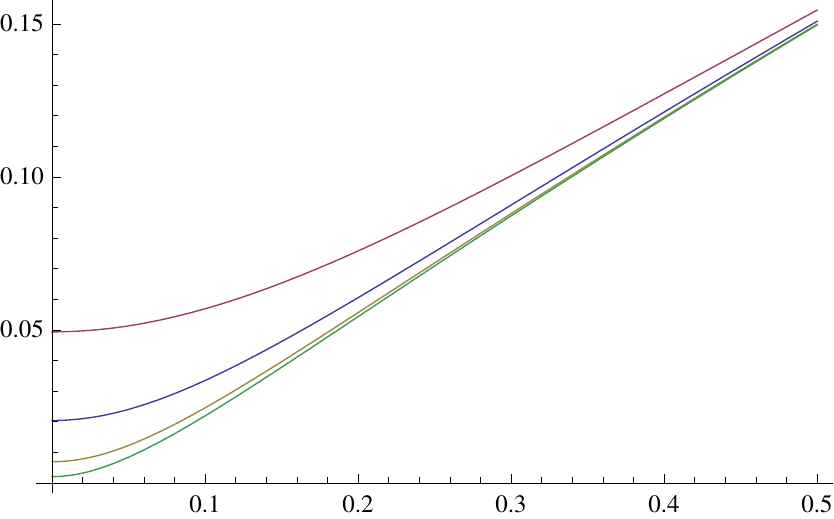}\
\includegraphics[width=5.1cm]{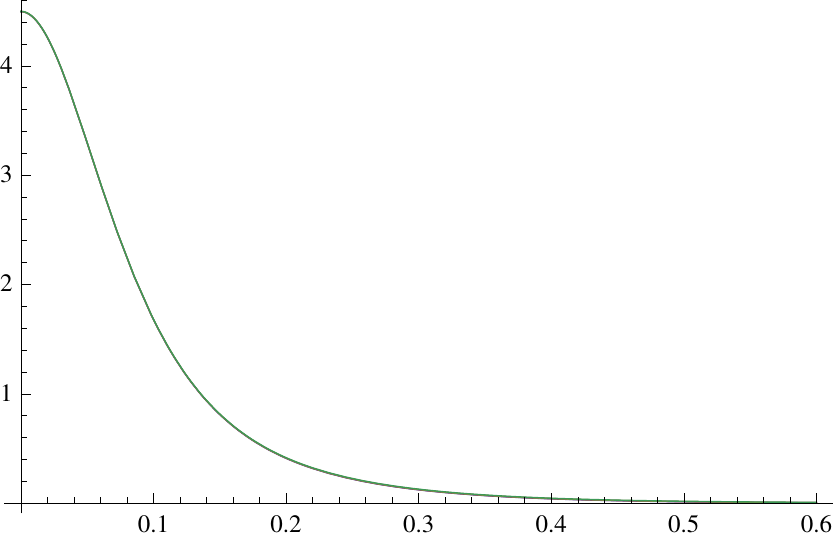}\
\includegraphics[width=6.2cm]{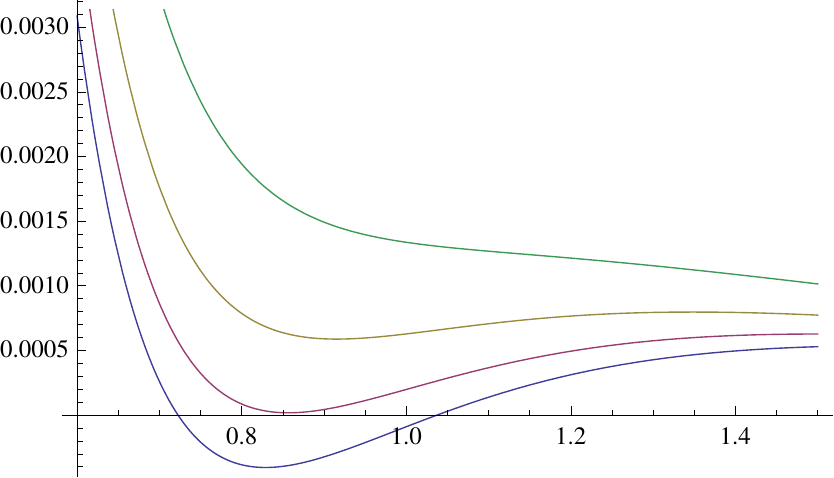}
\caption{Plots for Example QBH2.
	{\bf Left:} the metric function $\e^{2\gamma (x)}$ for $m=1$, $a=0.5$
	and $c = 0.025,\ 0.5,\ 0.1,\ 0.2$ (bottom-up).
	{\bf Middle:} the density $\rho_m(x)$ at small $x$ for $m=1$, $a=0.5$,
	$c = 0.1$ and $b$ ranging from 0.3 to 0.4: the $b$ dependence is
	indistinguishable.
	{\bf Right:} the same for $b = 0.39,\ 0.395,\ 0.398,\ 0.4$ at
	larger values of $x$ where the $b$ dependence is significant
	(upside down).
	  }
\end{figure}
  For instance, we can assume, as in (\ref{A3}),
\beq                                                        \label{chi4}
	\chi' = b/y^2, \cm b = \const > 0
\eeq
  with sufficiently small $b$. Examples of the behaviors of $\e^{2\gamma}$
  and $\rho_m$ close enough to the center for certain model parameters are
  presented in Fig.\,2. The left plot illustrates the proximity of
  $\e^{2\gamma}$ to zero near the center. The right plot makes it clear
  that, with the chosen values of $m,\,a,\,c$, we obtain $\rho_m > 0$ if $b
  < b_0 \approx 0.398$.

  The expressions for the electric and scalar charge densities are again too
  large to be reproduced here; it is important that they are finite and
  regular, but their particular form can add nothing to our understanding of
  the situation.

\section{Cylindrical symmetry}

  Besides starlike structures described as \sph\ configurations,
  of certain interest can be stringlike ones well described in the framework
  of \cy\ symmetry. The corresponding static metric of the form (\ref{ds})
  reads
\beq                                                          \label{ds2}
	ds^2 = \e^{2\gamma} dt^2
			- \e^{-2\gamma}(dx^2 + dz^2 + x^2 d\varphi^2),
\eeq
  where $z\in\R$ and $\varphi \in [0,2\pi)$ are the longitudial and
  azimuthal coordinates. All unknowns are functions of $x$.
  The field equations are similar to (\ref{chi1})--(\ref{equi1}):
\bear                                                        \label{chi2}
	2\eps x^{-1}\e^{2\gamma}(x\chi')'
			+ P_\chi \phi'{}^2 \eql - 8\pi J(x),
\\                                                           \label{phi2}
	x^{-1} \big(P \e^{-2\gamma} x\phi'\big)'
			\eql 4\pi \rho_e \e^{-3\gamma},
\\                                                           \label{gam2}
	x^{-1} \e^{2\gamma} (x\gamma')' - P \phi'{}^2
			\eql 4\pi \rho_m,
\\                                                           \label{int2}
	\gamma'{}^2 + \eps \chi'{}^2 \eql \e^{-2\gamma} P \phi'{}^2,
\\                                                           \label{equi2}
	\rho_m \gamma' - \rho_e \phi' \e^{-\gamma} \eql J\chi',
\ear
  the only change is that $x^2$ is replaced by $x$ at some places;
  instead of (\ref{rho_m1}) we obtain
\beq                                                          \label{rho_m2}
	4\pi \rho_m = \e^{2\gamma} (\gamma'' + \gamma'/x
			- \gamma'{}^2 -\eps\chi'{}^2).
\eeq
  These small changes in the equations, however, drastically affect the
  resulting physical picture. The main features of possible nonsingular \cy\
  solutions can be described as follows.

\medskip\noi
{\bf A regular axis} is obtained in the metric (\ref{ds1}) at $x=0$
  if and only if at small $x$
\beq                                                         \label{gam-ax}
  	\gamma(x) = \gamma_a + O(x^2), \cm \gamma_a = \const.
\eeq
  As near a regular center, we can use the Taylor expansion (\ref{A-c}),
  and it again follows that $\rho_m > 0$
  near the axis as long as $g_{00} =  A(x)$ has a minimum there.

\medskip\noi
{\bf Horizons.}
  A horizon at a finite \cy\ radius $r = x\e^{-\gamma}$ can exist at $x=0$
  and, as in \sph\ systems, it is necessarily extremal (double). However, its
  geometry, being regular, is still quite peculiar: the longitudinal metric
  coefficient $g_{zz}$ blows up there, so the horizon area is infinite, just
  as it happens in the so-called cold \bhs\ \cite{cold, cold-q}.

  Using the same Taylor expansion (\ref{A-hor}) as before, with $A_2 > 0$,
  we now obtain quite a different result for matter density near the
  horizon (assuming a finite value of $\chi'(0)$):
\beq
	4\pi \rho_m = - \sqrt{2/A_2} + O(x^2).
\eeq
  Thus the mass density is always negative at a horizon. It is a general
  result for \cy\ systems in our framework.

\medskip\noi
{\bf Asymptotic flatness.} This property, though desirable if a stringlike
  configuration should be observable in our Universe, is rather rare in
  solutions to the field equations (recall that even the Levy-Civita
  well-known vacuum solution is \asflat\ only in the trivial case where it
  is simply flat). For an \asflat\ configuration with the metric (\ref{ds2})
  one can require, without loss of generality, $\e^\gamma \to 1$ as
  $x \to \infty$.\footnote
  	{Stringlike configurations with an angular defect would also be
	 admissible, but they are impossible with the metric (\ref{ds2}).}
  Furthermore, if we assume, at large $x$,
  $\e^{\gamma} \approx 1 + ax^{-n}$ with certain constants $a$ and $n > 0$,
  calculating the first three terms in (\ref{rho_m2}), we find that they
  make a positive contribution to $\rho_m$ (which is necessary for having a
  phantom-free configuration) if and only if $a > 0$.

  Now, suppose we have a regular \asflat\ configuration (e.g., something
  like a QBH) with a regular axis. Then, at large $x$ the function
  $A(x) = e^{2\gamma}$ is decreasing whereas at $x=0$ it has a minimum. It
  means that $A(x)$ necessarily has a maximum at some finite $x = x_0$.
  Putting in its neighborhood $A = A_0 + \half A_2 (x-x_0)^2 + \ldots$,
  with $A_2 < 0$ to provide a maximum, and substituting this expansion to
  (\ref{rho_m2}), we obtain that the first three terms in (\ref{rho_m2})
  contribute negatively to $\rho_m$ near $x_0$, so that $\rho_m > 0$ can be
  obtained only with $\eps = -1$.

  We conclude that neither \cy\ BHs nor stringlike DMP configurations with a
  regular axis (in particular, stringlike QBHs) can be phantom-free.
  Therefore we abstain from giving particular examples here.

\section {Conclusion}

  The well-known static MP systems in GR include gravitational and
  electromagnetic fields and electrically charged dust with equal densities
  $\rho_m = |\rho_e|$.  We have considered a generalization of MP systems,
  including a dilatonic scalar field (the dilatonic MP, or DMP systems) with
  an arbitrary coupling function $P(\chi)$ in the Lagrangian (\ref{L}),
  where the scalar field $\chi$ may be normal or phantom. As in MP systems,
  the metric has a conformally flat spatial part. A DMP system is
  characterized by the metric function $\gamma(x^i)$, the electric potential
  $\varphi(x^i)$, the dilaton field $\chi(x^i)$, and three densities, those
  of mass, $\rho_m(x^i)$, electric charge, $\rho_e(x^i)$, and scalar
  charge, $J(x^i)$. Let us enumerate the main results obtained.

\begin{enumerate}
\item   
	It has been shown that static configurations are possible with
	arbitrary functions $g_{00} = \e^{2\gamma(x^i)}$ ($i=1,2,3$) and
	$\chi = \chi(\gamma)$, for any regular coupling function $P(\chi)$,
	without any assumption of spatial symmetry.
\item     
	For general static systems, the field equations imply that the
	functions $\gamma (x^i)$, $\chi(x^i)$ and $\phi(x^i)$ are related,
	so that if, say, $\gamma(x^i) \ne \const$, then $\chi =\chi(\gamma)$
	and $\phi = \phi(\gamma)$. It is thus unnecessary to postulate
	the existence of such functional relations, as is often done.
\item       
	There are purely scalar analogs of MP systems, but only with
	phantom scalar fields. However, the corresponding point sources
	are different: an extreme BH for MP, a singularity for scalar MP.
\item         
	It has been shown that sufficiently general configurations with
	nontrivial scalar fields cannot be obtained without a nonzero scalar
	charge density $J$; this, however, does not forbid the existence of
	special solutions with $J \equiv 0$ for particular $P(\chi)$.
\item           
	There is a universal balance condition, (\ref{charges}), between the
	Schwarzschild mass and the electric and scalar charges, valid for
	any \asflat\ DMP systems, including those with horizons and/or
	singularities. It generalizes the results previously obtained for
	special cases (see, e.g., \cite{clem2}).
\item             
	In the case of spherical symmetry, the existence conditions have
	been formulated for BH and quasi-BH (QBH) configurations with smooth
	matter, electric charge and scalar charge density distributions.
	It turns out that horizons in DMP systems are second-order (extremal),
	in agreement with the general properties of QBHs \cite{lemos4}.
\item               
	For QBHs containing a small parameter $c$ whose nonzero value
	distinguishes them from BHs, different limiting transitions $c\to 0$
	are analyzed. They lead to universal solutions independent of the
	particular choice of a QBH configuration. The limiting metrics
	coincide with those obtained previously for MP systems.
\item                 
	Examples of \sph\ BH and QBH solutions have been obtained. Among them
	are phantom-free ones, that is, the mass density and the energy
	densities of both scalar and electromagnetic fields are nonnegative.
\item                   
	For cylindrically symmetric configurations, the conditions at a
	regular center, a possible horizon and at flat infinity have been
	formulated. It has been shown that neither \cy\ BHs nor stringlike
	DMP configurations with a regular axis (in particular, stringlike
	QBHs) can be phantom-free.
\end{enumerate}

  Some of these results have been briefly presented in \cite{we-14}, viz.,
  items 1, 3, 5, and 8 (partly). In addition, in \cite{we-14}, polycentric
  configurations, possible in the DMP framework, were discussed, with any
  number of mass concentrations. For instance, one can consider the metric
  (\ref{ds}) in Cartesian coordinates $x^i=(x,y,z)$ (so that $h_{ik}=
  \delta_{ik}$) and choose
\beq                                                         \label{f_n}
	\e^{-\gamma(x^i)}\equiv  f(x^i)
		= \frac{1}{n} \sum_{a=1}^{n} f_a (X_a),
\eeq
  where $f_a$ are functions of $X_a := |x^i - x^i_a|$, $x^i_a$ being the
  (fixed) coordinates of the $a$-th center. As $f_a$, one can take any
  functions providing \asflat\ \sph\ solutions, e.g., BHs or QBHs.
  A complete solution is obtained after choosing the function
  $\chi(\gamma)$, or equivalently $\chi(f)$, which should be regular at all
  relevant values of $f$ and decay sufficiently rapidly at spatial infinity,
  as $f \to 1$. In \cite{we-14}, an example is given of such a system
  with two mass concentrations, where each ``center'' can be a BH or a QBH.

  We would like to stress that it is in general rather difficult to find
  sources that admit QBH configurations since matter begins to collapse long
  before approaching a would-be horizon. It is the freedom in choosing
  the metric function $\gamma(x)$ that enables us to keep DMP configurations
  static even extremely closely to emergence of a horizon.

\subsection*{Acknowledgments}

  We thank CNPq (Brazil) and FAPES (Brazil) for partial financial support.


\small
\begin{thebibliography}{10} 

\bibitem{h-e}
	S.W. Hawking and G.F.R. Ellis, {\it The Large Scale Structure of
	Space-Time} (Cambridge University Press, Cambridge, 1973).

\bibitem{fro-nov}
	V.P. Frolov and I.D. Novikov, {\it Black Hole Physics.
	Basic Concepts and New Developments } (Kluwer, 1997).

\bibitem{majumdar}
	S.D. Majumdar, Phys. Rev. {\bf 72}, 390 (1947).

\bibitem{papa}
	A. Papapetrou, Proc. Roy. Irish Acad. {\bf A51}, 191 (1947).

\bibitem{lemos1}
        Jos\'e P.S. Lemos and E.J. Weinberg, Phys. Rev. {\bf D 69}, 104004
	(2004).

\bibitem{qbh}
	Jos\'e P.S. Lemos and O.B. Zaslavskii, Phys. Rev. D \textbf{76},
	084030 (2007).

\bibitem{lemos2}
	Jos\'e P.S. Lemos and O.B. Zaslavskii, Phys. Rev. {\bf D 82},
	024029 (2010).

\bibitem{lemos3}
        Jos\'e P.S. Lemos and O.B. Zaslavskii, Phys. Lett. {\bf B 695}, 37
	(2011).

\bibitem{lemos4}
	Jos\'e P.S. Lemos,
	Scientific Proceedings of Kazan State University
 	(Uchonye Zapiski Kazanskogo Universiteta (UZKGU)) {\bf 153}, 215
	(2011); arXiv: 1112.5763.

\bibitem{bon99}
	W.B. Bonnor, Class. Quantum Grav. {\bf 16}, 4125 (1999).

\bibitem{bon10}
	W.B. Bonnor,
	Gen. Rel. Grav. {\bf 42}, 1825 (2010).

\bibitem{mein11}    
	R. Meinel and M. H\"utten,
	Class. Quantum Grav. {\bf 28}, 225010 (2011).

\bibitem{fish}
	I.Z. Fisher, Zh. Eksp. Teor. Fiz. {\bf 18}, 636 (1948); gr-qc/9911008.

\bibitem{pha-57}
	O. Bergmann and R. Leipnik, Phys. Rev. {\bf 107}, 1157 (1957).

\bibitem{br73}
	K.A. Bronnikov, Acta Phys. Pol. {\bf B4}, 251 (1973).

\bibitem{boch}
	N.M. Bocharova, K.A. Bronnikov and V.N. Melnikov,
        Vestnik MGU, Fiz., Astron., No.6, 706 (1970).

\bibitem{BD}
	C. Brans and R.H. Dicke, Phys. Rev. {\bf 124}, 925 (1961).

\bibitem{cold1}
	K.A. Bronnikov, G. Cl\'ement, C.P. Constantinidis and J.C. Fabris,
	Phys. Lett. {\bf A243}, 121 (1998), gr-qc/9801050.
	
\bibitem{cold2}
	K.A. Bronnikov, G. Cl\'ement, C.P. Constantinidis and J.C. Fabris,
	Grav. Cosmol. {\bf 4}, 128 (1998), gr-qc/9804064.

\bibitem{cold-q}
	K.A. Bronnikov, C.P. Constantinidis, R.L. Evangelista and J.C. Fabris.
        Int. J. Mod. Phys. D {\bf 8}, 481 (1999); gr-qc/9902050.

\bibitem{dam1}
	T. Damour and A.M. Polyakov,
	Nucl. Phys. B {\bf 423}, 532--558 (1994); hep-th/9401069.

\bibitem{dam2}
	T. Damour and A.M. Polyakov,
	Gen. Rel. Grav. 26, 1171--1176 (1994); gr-qc/9411069.

\bibitem{BSh-77}
	K.A. Bronnikov and G.N. Shikin,
	Izvestiya Vuzov SSSR, Fiz., No. 9, 25 (1977).

\bibitem{BSh-77e}
	K.A. Bronnikov and G.N. Shikin,
        Russ. Phys. J. {\bf 20} (9), 1138--1143 (1977).

\bibitem{dbh1}
	G. W. Gibbons and K. Maeda, Nucl. Phys. B {\bf 298}, 741 (1988).

\bibitem{dbh2}
	D. Garfinkle, G.T. Horowitz, and A. Strominger,
	Phys. Rev. D {\bf 43}, 3140 (1991);
	Erratum: Phys. Rev. D {\bf 45}, 3888 (1992).

\bibitem{dbh3}
	M. Rakhmanov, Phys. Rev. D {\bf 50}, 5155 (1994)

\bibitem{BMsh-78}
	K.A. Bronnikov, V.N. Melnikov, and G.N. Shikin,
	Izvestiya Vuzov SSSR, Fiz., No. 11, 69 (1978).

\bibitem{BMsh-78e}
	K.A. Bronnikov, V.N. Melnikov, and G.N. Shikin,
	Russ. Phys. J. {\bf 21} (11), 1443 (1978).

\bibitem{ann-79}
	K.A. Bronnikov, V.N. Melnikov, G.N. Shikin, and K.P. Staniukovich,
	Ann. Phys. (NY) {\bf 118} (1), 84 (1979).

\bibitem{bra-r1}
    	V.D. Ivashchuk and V.N. Melnikov,
	Class. Quantum Grav. {\bf 14}, 3001 (1997);
        Corrigendum ibid. {\bf 15}, 3941 (1998); hep-th/9705036.

\bibitem{bra-r2}
	V.D. Ivashchuk and V. N. Melnikov, Class. Quantum Grav.
        {\bf 18}, R87--R152 (2001); hep-th/0110274.

\bibitem{kb95}
	K.A. Bronnikov, \GC {1} 67--78 (1995).

\bibitem{bra1}
	K.A. Bronnikov, V.D. Ivashchuk and V.N. Melnikov,
        Grav. Cosmol. {\bf 3}, 203--212 (1997).

\bibitem{bra2}
	K.A. Bronnikov,
        Grav. Cosmol. {\bf 4}, 49--56 (1998); hep-th/9710207.

\bibitem{bra3}
	V.D. Ivashchuk and V. N. Melnikov, \GC {17} 328--334 (2011).

\bibitem{clem1}
	G. Cl\'ement, J.C. Fabris, and M.E. Rodrigues,
	Phys. Rev. {\bf D 79}, 064021 (2009).

\bibitem{clem2}
	M. Azreg-Ainou, G. Cl\'ement, J.C. Fabris, and M.E. Rodrigues,
	\PRD {83} 124001 (2011); arXiv: 1102.4093.

\bibitem{yaz}
	S. Yazadjiev, Mod. Phys. Lett. A {\bf 20}, 821 (2005);
	arXiv: gr-qc/0411132.

\bibitem{we-14}
	K.A. Bronnikov, J.C. Fabris, R. Silveira, and O.B. Zaslavskii,
	Phys. Rev. D {\bf 89}, 107501 (2014); ArXiv: 1405.6116.
\bibitem{BR-book}
	K.A. Bronnikov and S.G. Rubin,
	{\it Black Holes, Cosmology, and Extra Dimensions }
     	(World Scientific, 2012). 


\bibitem{ger}
	R. Geroch, Commun. Math. Phys. \textbf{13}, 180 (1969).

\bibitem{br1}
	B. Bertotti, Phys. Rev. \textbf{116}, 1331 (1959).

\bibitem{br2}
	I. Robinson, Bull. Acad. Pol. Sci. \textbf{7}, 351 (1959).

\bibitem{an1}
	F.G. Alvarenga, A.B. Batista, J.C. Fabris, and G.T. Marques,
	Phys. Lett. A {\bf 320}, 83 (2003).

\bibitem{an2}
	F.G. Alvarenga, A.B. Batista, J.C. Fabris, and G.T. Marques,
	Grav. Cosmol. {\bf 10}, 184 (2004).

\end{thebibliography}
\end{document}